\documentclass[prl,reprint,
nolongbibliography, 
aps,amsmath,amssymb,
]{revtex4-2}
\usepackage{graphicx}% Include figure files
\usepackage{xcolor}
\usepackage{braket}
\usepackage{bm}% bold math
\usepackage{hyperref}% add hypertext capabilities
\usepackage{ragged2e}  % 左詰め用
\begin{document}

%\preprint{APS/123-QED}

\title{Holographic AME states in black hole interiors}% Force line breaks with \\
%\thanks{A footnote to the article title}%

\author{Takanori Anegawa}
 \email{takanegawa@gmail.com}
 \affiliation{Yonago College, National Institute of Technology
Yonago, Tottori 683-8502, Japan}%Lines break automatically or can be forced with \\
\author{Kotaro Tamaoka}%
 \email{tamaoka.kotaro@nihon-u.ac.jp}
\affiliation{%
Department of Physics, College of Humanities and Sciences, Nihon University, Sakura-josui, Tokyo 156-8550}%

\date{\today}% It is always \today, today,
             %  but any date may be explicitly specified

\begin{abstract}
We argue that the special extremal slice inside an AdS black hole is dual to an absolutely maximally entangled (AME) state. We demonstrate this by confirming the $n$-independence of holographic $n$-th Renyi entropies for any bi-partite subsystems. Our result gives an AME state in an infinite-volume system, where the local bond dimension is set by the black hole entropy. In particular, our construction provides concrete support from the gravity side for the emergence of random structures and an infinite-dimensional Hilbert space in recent non-isometric holographic codes.
\end{abstract}

%\keywords{Suggested keywords}%Use showkeys class option if keyword
                              %display desired
\maketitle

%\tableofcontents
%\onecolumngrid 

\section{Introduction and Summary}
\begin{figure}[t]
    \centering
\includegraphics[width=0.45\columnwidth]{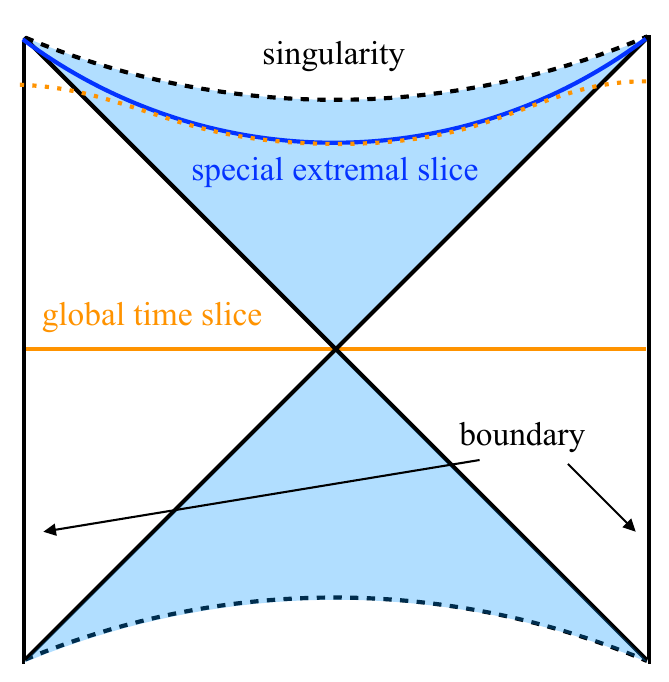} 
    \caption{\justifying A time slice (blue solid line in shaded region) called special extremal slice in eternal AdS black hole. We view it as a holographic screen on which dual theory is defined. Orange line corresponds to the time slice dual to TFD state in CFTs. A time evolution of TFD state at the late time gives the extremal surface (orange dotted line) that has partial contact with special extremal slice. }
    \label{fig:bh1}
\end{figure}

Recent advances in quantum gravity are heavily based on the progress of the AdS/CFT correspondence~\cite{Maldacena:1997re}, a concrete realization of the holographic principle~\cite{tHooft:1993dmi, Susskind:1994vu}. In its conventional formulation, the holographic screen is placed at asymptotic infinity, where the dual CFT is defined on the same manifold. From this perspective, describing the interior of black holes is highly nontrivial, as the theory is formulated in terms of boundary-time coordinates. To address this, Maldacena proposed that the global time slice of maximally extended AdS-Schwarzschild black hole is dual to an entangled state between two CFTs, 
\begin{align}
\text{global time slice}\leftrightarrow\ket{\text{TFD}_\beta}
\end{align}
where,
\begin{align}
\ket{\text{TFD}_\beta}=Z(\beta)^{-\frac{1}{2}}\sum_ne^{-\frac{\beta}{2}E_n}\ket{n_Ln_R}.
\end{align}
known as the thermofield double (TFD) state (see FIG.\ref{fig:bh1})~\cite{Maldacena:2001kr}. While we obtain a CFT description of the black hole interior, decoding the interior information from the CFT remains indirect and complex~\cite{Fidkowski:2003nf, Harlow:2013tf,Hayden:2007cs, Brown:2019rox}, reflecting the formulation in terms of exterior coordinates of the black hole. %Indeed, understanding the black hole interior remains one of the central challenges in holography. 

Given the equivalence principle, it is natural to expect a more direct and simpler description of the interior, at least at the semiclassical level. In this paper, we address this problem by considering holographic screens placed directly in the black hole interior. For recent approaches that utilize $T\overline{T}$ deformations~\cite{Zamolodchikov:2004ce,Smirnov:2016lqw,McGough:2016lol} and their variants to realize such screens, see \cite{Araujo-Regado:2022gvw,Soni:2024aop,AliAhmad:2025kki}, for example.

We show that the quantum state dual to a particular time slice in the black hole interior, called the special extremal slice~\cite{Hartman:2013qma}, takes a remarkably simple form: a maximally entangled state. More precisely, we argue that this state is an example of an absolutely maximally entangled (AME) state~\cite{Helwig:2012nha,Helwig:2013qoq}, 
\begin{align}
\text{special extremal slice} &\leftrightarrow \ket{\text{AME}}, \label{eq:ses-ame}
\end{align}
which is maximally entangled across all bipartitions.
AME states play an important role especially in quantum information science, serving as key resources for quantum error correction and secret sharing~\cite{Helwig:2012nha}.
These states have also been used as theoretical tools in AdS / CFT toy models based on quantum error-correcting codes~\cite{Almheiri:2014lwa,Pastawski:2015qua}. Inspired by the appearance of AME state in such models, the similar notion called fixed area state~\cite{Dong:2018seb} has been proposed and has proven useful for various computations in gravitational path integral. However, it has remained unclear whether an AME state can, by itself, represent a classical spacetime. To our knowledge, this provides the first example in AdS/CFT where a single AME state, without any superposition, directly describes a time slice of classical spacetime, most notably in the interior.

We support our proposal \eqref{eq:ses-ame} through a holographic calculation of Renyi entropy~\cite{Dong:2016fnf},
\begin{align}
S^{(n)}_A=\dfrac{1}{1-n}\log\textrm{Tr}\rho^n_A,
\end{align}
where $\rho_A$ is the reduced density matrix for a subsystem $A$. In particular, $n\rightarrow1$ limit yields the entanglement entropy $S_A=-\textrm{Tr}\rho_A\log\rho_A$. Our analysis shows that the Renyi entropy $S^{(n)}_A$ is independent of $n$, {\it i.e.} $S^{(n)}_A=S_A$, confirming that the original state is maximally entangled. 

In our setting, AME state can also be well approximated by a Haar-random state,
\begin{align}
\ket{\text{AME}}\simeq U\ket{\psi_0},\label{eq:appHaar}
\end{align}
where $U$ is a unitary operator drawn from the Haar measure and $\ket{\psi_0}$ is an arbitrary reference state. The local Hilbert space dimension is set by the density of the black hole entropy, and the approximation becomes exact in the semiclassical limit. While the emergence of such randomness has long been recognized in the exterior dynamics of black holes~\cite{Page:1993wv,Page:1993df}, our results demonstrate that it naturally appears in the interior as well. Our findings also justify the use of Haar-random structures in recent non-isometric code models~\cite{Akers:2022qdl} that aim to capture the structure of the Hilbert space in the black hole interior. Furthermore, our results also naturally explain a key feature of such models: the presence of infinitely many interior states (called null states) exceeding the black hole entropy. In our setting, the existence of these states follows directly from the fact that the timelike direction for exterior becomes spacelike inside the black hole.

\section{Setup and Review}

We consider static black holes in AdS spacetime, including both the BTZ black hole~\cite{Banados:1992wn} in global coordinates and higher-dimensional AdS-Schwarzschild black holes expressed in Poincare coordinates. While the conventional AdS/CFT correspondence introduces a UV cutoff surface at large radius $ r = r_b \gg 1$ in global coordinates, or equivalently at small $z = z_b \ll 1 $ in Poincare coordinates, we instead consider the surfaces located inside the black hole horizon as FIG.\ref{fig:bh1}.

It is important to note that, in the black hole interior, the radial coordinates $ r $ or $ z $ play the role of time. Thus, constant-$ r $ or constant-$ z $ slices correspond to a time slice. We assume the existence of a dual quantum theory defined on such a time slice. We emphasize that this theory is formulated in a Euclidean signature. In this sense, our construction resembles aspects of the dS/CFT correspondence~\cite{Strominger:2001pn,Witten:2001kn,Maldacena:2002vr}, where dual theories are also defined on the same manifold. Although we do not assume any particular construction, this is a setting similar to that of Cauchy slice holography in \cite{Araujo-Regado:2022gvw}. We will discuss these points further in the discussion section. As bulk wave functions in these set up, we will focus on the Diriclet boundary condition for the induced metric on the constant $r$ slice. 

A further important point is that the asymptotic boundary time coordinate $t$, which would normally be used in boundary formulations, becomes spacelike in the black hole interior. We consider subsystems that extend along this $ t $-direction and evaluate entanglement and Renyi entropies holographically. Since the setup involves a time-dependent background, the corresponding extremal surfaces~\cite{Hubeny:2007xt} must be evaluated, rather than the minimal surfaces~\cite{Ryu:2006bv}. This prescription applies to boundaries other than the asymptotic infinity when the semiclassical limit is well defined~\cite{Lewkowycz:2013nqa}, allowing our result to be identified with the entropies of the dual theory, if it exists.

For holographic analysis of Renyi entropy, it is convenient to introduce a refined version,
\begin{align}
\tilde{S}^{(n)}_A=n^2\partial_n\left(\dfrac{n-1}{n}S^{(n)}_A\right),
\end{align}
which is known to have a gravity dual~\cite{Dong:2016fnf} given by
\begin{align}
\tilde{S}^{(n)}_A=\dfrac{A(C_n)}{4G_N},
\end{align}
where $A(C_n)$ is the area of a codimension-2 cosmic brane $C_n$ with tension
\begin{align}
T_n=\dfrac{n-1}{4nG_N}.
\end{align}
For general $n>1$, the cosmic brane causes the backreaction onto the original geometry, creating a conical deficit angle
\begin{align}
\Delta\phi=2\pi\left(1-\dfrac{1}{n}\right)
\end{align}
for the ambient geometry. Taking $n\rightarrow1$, we obtain entanglement entropy and corresponding area of extremal surface where there is no back-reaction onto the original spacetime. 

Finally, we recall the original setup where the special extremal slices naturally appear~\cite{Hartman:2013qma}.  These slices appear when we holographically estimate the entanglement entropy between two CFTs in the TFD state dual to the maximally extended AdS black hole. There, the entanglement entropy of large intervals at late times is computed by evaluating extremal surfaces that pass through a distinguished interior time slice, known as the special extremal slice (see FIG.\ref{fig:bh1}). More specifically, this slice is defined as the point where a certain function $a(\kappa)$ (introduced later) gives its extremum, which is in fact a maximum. In our construction, we identify this special slice with the holographic screen on which the dual theory is defined~\footnote{This slice generally differs from the maximal-volume slice used in holographic complexity, as the maximization is performed over different codimension objects.}. 

\section{BTZ Black Hole}

We begin our analysis with the static BTZ black hole, described by the metric
\begin{align}
ds^2 = (r_+^2 - r^2)\, dt^2 - \frac{dr^2}{r_+^2 - r^2} + r^2\, d\phi^2. \label{eq:BTZ}
\end{align}
For $r < r_+$, the $t$-direction becomes spacelike, while $r$ behaves as a timelike coordinate. We place our holographic screen at $r = r_b < r_+$ and impose Dirichlet boundary conditions on the induced metric at this surface. Of particular interest is the limit $r_b \to 0$, which corresponds to the special extremal slice introduced in \cite{Hartman:2013qma}.

In this limit, the entanglement entropy associated with spatial subsystems along the $\phi$-direction becomes trivial, similar to the situation in static patch holography in de Sitter space~\cite{Susskind:2021omt}. A key difference, however, is that in the present context, intervals along the $t$-direction (e.g., $[-T/2, T/2]$) become spacelike and can serve as entangling regions.

We first evaluate the holographic entanglement entropy. Since the background is time-dependent, extremal (rather than minimal) surfaces must be considered. Solving the geodesic equations with boundary conditions $(r(\pm\lambda_*), \phi(\pm\lambda_*), t(\pm\lambda_*)) = (0, \phi_0, \pm T/2)$, we obtain a straight line located at
\begin{align}
r(\lambda) = 0 \quad (-\lambda_* \leq \lambda \leq \lambda_*),
\end{align}
with $\phi(\lambda) = \phi_0$, $t(\lambda) = \lambda / (2r_+)$, and $\lambda_* = r_+ T$. Here, $\lambda$ is the proper length parameter. Since the extrinsic curvature vanishes at $r=0$, the boundary geodesics also solve the bulk geodesic equations. The area of the extremal surface is thus $2\lambda_* = r_+ T$, leading to the holographic entanglement entropy
\begin{align}
S_A = \frac{r_+}{4G_N} T = \frac{S_{\text{BH}}}{2\pi} T,
\end{align}
where $S_{\text{BH}} = 2\pi r_+ / (4G_N)$ is the Bekenstein-Hawking entropy of the BTZ black hole. The quantity $S_{\text{BH}} / 2\pi$ can be interpreted as the entropy density of the dual theory on the conventional asymptotic boundary.

This result implies a volume law for entanglement in the boundary theory. The entropy can easily exceed $S_{\text{BH}}$ as the subsystem size increases, which is natural given that the dual theory is defined on an infinite-volume slice along the $t$-direction. There is no UV divergence, suggesting that the boundary theory is a nonstandard quantum theory, such as $T\overline{T}$-deformed theory. %It is straightforward to extend this computation to multiple disjoint intervals, yielding the same form of entropy with $T$ replaced by the total length of the entangling region.

We now consider the refined Renyi entropy. Since we are working in three-dimensional Einstein gravity, the backreacted geometry outside the brane remains locally AdS$_3$. Moreover, the Dirichlet condition at $r=0$ completely fixes the bulk geometry to be the original BTZ metric \eqref{eq:BTZ}, even after inserting a codimension-2 brane with tension. In this setup, the brane remains located at $r = 0$ along the $t$-direction, and the resulting deficit angle appears along the $\phi$-direction. This backreaction does not affect the length of the brane itself. As a result, the holographic refined Renyi entropy is given by
\begin{align}
\tilde{S}_A^{(n)} = \frac{r_+}{4G_N} T.
\end{align}
We therefore conclude that the $n$-dependence drops out:
\begin{align}
S_A^{(n)} = S_A.
\end{align}
That is, the entire wavefunction is a maximally entangled state. One might concern that the global state may be mixed. Since the area of the neck at $r=r_+$ is zero as $g_{tt}|_{r=r_+}=0$, the global state is indeed a pure state. 

These results can be easily generalized to arbitrary bipartitions of the Hilbert space. For example, if we consider two disjoint (finite) interval, $A=[t_1,t_2]$ and $B=[t_3,t_4]$, we always have
\begin{align}
S_{AB}^{(n)}=S^{(n)}_A+S^{(n)}_B=S_A+S_B,
\end{align}
as the geodesics are just straight lines. By contrast, in the same setup, taking $t_1 \to -\infty$ and $t_4 \to \infty$ leads to
\begin{align}
S_{AB}^{(n)} = \dfrac{S_{BH}}{2\pi}(t_3 - t_2) = S_{AB}.
\end{align}
Namely, we may obtain a non-trivial ``entanglement wedge'' only when the subsystem is infinite volume. 
In any case, we have $n$-independence of Renyi entropy, confirming that the state is an AME state.

\section{Higher dimensional genealization}

%\begin{figure}[t]  % "t" places the figure at the top of the page
%    \centering
%\includegraphics[width=0.6\columnwidth]{bh_fig2.pdf}  % Adjust width as needed
%    \caption{\justifying A schematic picture of the calculation of holographic entanglement entropy when a holographic screen is placed on the special extremal slice. Since the minimal surface does not extend beyond the slice, the surface coincides with the subsystem.}
%    \label{fig:bh2}
%\end{figure}

Next, we generalize our results for BTZ black hole to higher-dimensional static black holes in AdS. For simplicity, let us consider the planar black branes,
\begin{align}
    ds^2=\dfrac{1}{z^2}\left(-f(z)dt^2+\dfrac{dz^2}{f(z)}+d\vec{x}^2_{d-1}\right),
\end{align}
where 
\begin{align}
    f(z)=1-\dfrac{z^d}{z_H^d}.
\end{align}
Note that $d=2$ case is equivalent to the BTZ black hole with infinite volume limit (along the spatial direction). For later convenience, we apply the following coordinate transformation,
\begin{align}
z=z_H\left(\cosh\dfrac{d\rho}{2}\right)^{-\frac{2}{d}},
\end{align}
so that
\begin{align}
ds^2=-g(\rho)^2dt^2+d\rho^2+h(\rho)^2d\vec{x}_{d-1}^2,
\end{align}
where
\begin{align}
h(\rho)&=\dfrac{2}{d}\left(\cosh\dfrac{d\rho}{2}\right)^{2/d}, \\
g(\rho)&=h(\rho)\tanh\dfrac{d\rho}{2}.
\end{align}
For the $\rho$-coordinates, the horizon is at $\rho=0$. The interior region corresponds to pure imaginary value, $\rho=i\kappa$. The origin corresponds to $\kappa=\frac{\pi}{d}$. In particular, special extremal slice $\kappa_m$ is given by
\begin{align}
\tan\frac{d\kappa_m}{2}=\sqrt{\frac{d}{d-2}}.
\end{align}

Let us consider $x_{d-1}=0$ slice and take a subsystem $A$ specified by the coordinates $(t,x_1,\cdots,x_{d-2})$. We take boundary condition as $\kappa|_{\partial A}=\kappa_m$. The area functional is given by
\begin{align}
A=\int dtd^{d-2}x\;a(i\kappa)\sqrt{1-\frac{(\partial_t\kappa)^2}{|g(i\kappa)^2|}-\sum_{j=1}^{d-2}\frac{(\partial_j\kappa)^2}{h(i\kappa)^2}},
\end{align}
where $a(i\kappa)=|g(i\kappa)|h(i\kappa)^{d-2}$. Note that in the interior, the Euler-Lagrange equation becomes quasi-linear elliptic differential equation. Therefore, one can apply the strong maximum principle~\cite{Andersson1998} as like Laplace equation~\footnote{To use the theorem, we need to assume $\kappa$ to be $C^1$ for each variable and restrict $\kappa$ to $0<\kappa<\kappa_0<\pi/d$.}. It means that we have a unique solution,
\begin{align}
\kappa(t,\vec{x}_{d-2})=\kappa_m,
\end{align}
and obtain
\begin{align}
S_A=\dfrac{A}{4G_N}=\dfrac{a(i\kappa_m)}{4G}\min[V_A,V_{\bar{A}}],
\end{align}
where $V_A$ and $V_{\bar{A}}$ are size of subsystem $A$ and its complement $\bar{A}$. To be concrete and illustrate the idea, suppose we take a subsystem defined as a finite interval $t=[-T/2,T/2]$ times an infinite strip $(x_1, \cdots,x_{d-2})\in\mathbb{R}^{d-2}, x_{d-1}=0$. Then, the area functional reduces to 
\begin{align}
A=V_{d-2}^{\perp}\int dt [h(i\kappa)]^{d-2}\sqrt{-g^2(i\kappa)-\dot{\kappa}^2}
\end{align}
where $V_{d-2}^{\perp}$ is the transverse volume. Solving the first integral for $\dot{\kappa}$ gives
\begin{align}
\dot{\kappa}^2=\dfrac{|g(i\kappa)|^2}{E^2}\left(E^2-|g(i\kappa)|^2h^{2d-4}(i\kappa)\right).
\end{align}
Here $E$ is conserved energy from time translation invariant of the action and is determined from the boundary condition $\kappa(\pm T/2)=\kappa_m$ as
\begin{align}
E&=\begin{cases}
    1& (d=2),\\
   \frac{2^{-2+d}d^{1-d}\sqrt{d(d-2)}(1+\frac{d}{d-2})^{\frac{1}{d}}}{d-1}& (d>2).
\end{cases}
\end{align}
\begin{figure}[t]
    \centering
\includegraphics[width=0.7\columnwidth]{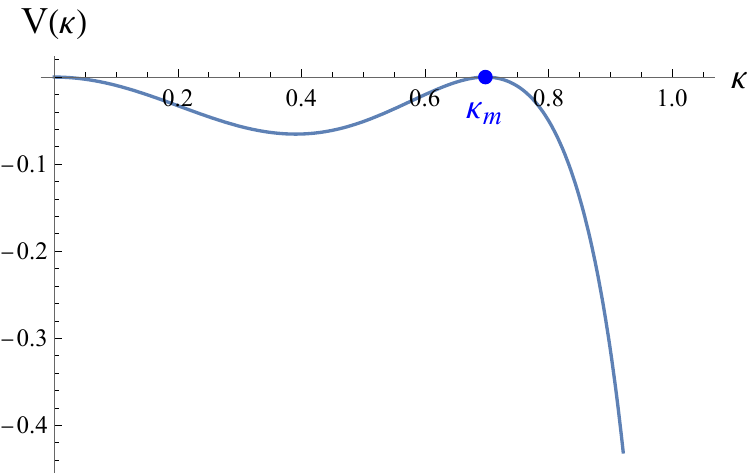} 
    \caption{\justifying Example of potential problem in $d=3$. There are two extrema (the horizon $\kappa=0$ and the special extremal slice $\kappa=\kappa_m$), and we start from $\kappa=\kappa_m$. Note that we now require that $\kappa = \kappa_m$ at both ends of the finite time interval. Therefore, only allowed solution that comes back to $\kappa=\kappa_m$ in finite time is $\kappa(t)=\kappa_m$. }
    \label{fig:d3}
\end{figure}
One can view this as a problem of classical mechanics, $\dot{\kappa}^2=-V(\kappa)$ (see FIG.~\ref{fig:d3}). It is easy to see that all allowed solutions that can back to $\kappa_m$ is $\kappa(t)=\kappa_m$. In this setup, we obtain
\begin{align}
    S_A=\dfrac{a(i\kappa_m)}{4G_N}V_{d-2}^\perp T,
\end{align}
If we take subsystem as entire system, the extremal surface is allowed to go to $\kappa(t)=0$, but then the area becomes zero due to the vanishing $g(\kappa)$ at $\kappa=0$ --- consistent with the nature of global pure state. 

Furthermore, these arguments readily generalize to the $n$-th Renyi entropy for any bipartition. Specifically, the unique constant solution $\kappa(t,\vec{x}_{d-2})=\kappa_m$ together with the Dirichlet boundary condition ensures that $S^{(n)}_A=S_A$. This reasoning closely parallels that for fixed-area states~\cite{Dong:2018seb}.

\section{Connection to non-isometric code}

We have shown that for \textit{any} bipartition the Renyi entropies satisfy $S_A^{(n)} = S_A$, establishing that the state on the special extremal slice is AME.  Treating the interior Hilbert space as a tensor product
$\mathcal H=\mathcal H_A\otimes\mathcal H_{\bar{A}}$ with large local dimensions
$d_{A,\bar{A}}$, the same state is well-approximated by a Haar-random vector as mentioned around Eq.\eqref{eq:appHaar}.  In particular,
\begin{equation}
\mathbb {E}_{U}\!\left[\operatorname{Tr}\rho_A^{\,n}\right]
      = \frac{1}{d_A^{n-1}}
        \left(1+\mathcal O\!\left(\frac{d_A}{d_{\bar{A}}}\right)\right),
\label{eq:Haar_average}
\end{equation}
whenever $d_A\!\ll\! d_{\bar{A}}$. Here $\rho_A=\textrm{Tr}_{\bar{A}}[U\ket{\psi_0}\!\bra{\psi_0}U^\dagger]$ and $\mathbb {E}_{U}$ is average of unitary operators by Haar measure. 
The assumption is justified by taking the semiclassical limit $G_N \to 0$ and noting that the $t$-direction has infinite length in the interior. Eq. \eqref{eq:Haar_average} therefore holds and one can approximate our AME state as Haar random state, $\ket{\text{AME}}\simeq U\ket{\psi_0}$. 

These observations agree with the emerging picture of the black hole interior as a non-isometric quantum error-correcting code.  In the usual (isometric) AdS/CFT map the low-energy AdS Hilbert space is embedded isometrically into the code subspace of the boundary CFT~\cite{Almheiri:2014lwa}.  By contrast, the interior Hilbert space is believed to be larger than the Hilbert space of CFT, forcing interior-to-boundary map to be non-isometric.  A toy model proposed in Ref.~\cite{Akers:2022qdl} takes the form
\begin{equation}
V \;=\; \sqrt{|P|}\,
        \langle 0|_{P}\,U ,
\label{eq:noniso}
\end{equation}
where $U$ is Haar random on an interior Hilbert space
$\mathcal H_{\text{int}}$ and post-selection by $|0\rangle_P$
land in the fundamental Hilbert space $\mathcal H_B$~\footnote{Here we suppress auxiliary qubits for interior Hilbert space in order to align degrees of freedom for input and output. These qubits are not important in the present context.}.
Our gravity construction provides direct support
for the two key assumptions behind~\eqref{eq:noniso}:
\begin{figure}[t]
    \centering
\includegraphics[width=0.7\columnwidth]{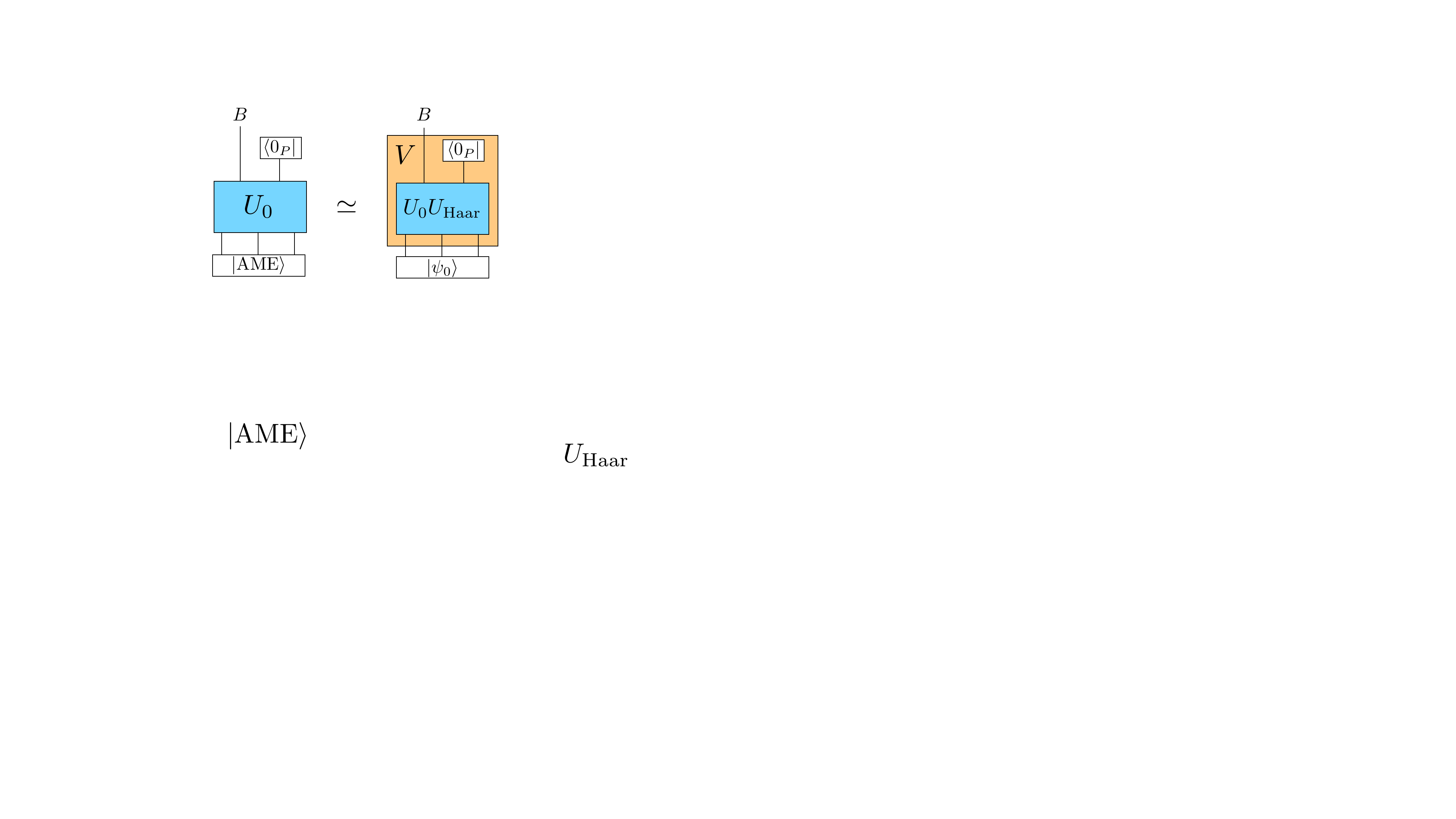} 
    \caption{\justifying On the state on special extremal slice, even if we start from a particular $U_0$ for the non-isometric code, we have natural replacement  $U_0\rightarrow U^\prime_{\text{Haar}}\equiv U_{0}U_{\text{Haar}}$.}
    \label{fig:haar}
\end{figure}
\begin{enumerate}
\item \emph{Haar randomness.}  
      The AME state on the special extremal slice is indistinguishable
      from a Haar-random vector, justifying the appearance of $U$ in~\eqref{eq:noniso}. See FIG.\ref{fig:haar}

\item \emph{Dimension of interior Hilbert space.}  
      Because the interior $t$-direction is spacelike and unbounded, the effective dimension $\dim\mathcal H_{\text{int}}$ is formally infinite~\footnote{The hierarchy was originally motivated in the entropy of evaporating black hole: if radiation is collected for a time interval $T>t_{\text{Page}}$, the radiation entropy grows linearly in time,
$S_R\propto T$, and can greatly exceed $S_{\text{BH}}=\log\dim\mathcal H_B$. 
In our framework, the interior entanglement entropy $S_A\propto T$ reproduces this growth, suggesting existence of the null states inside the black hole.}, whereas $\dim\mathcal H_B\sim e^{S_{\text{BH}}}$ is finite.  The inevitable many-to-one map $\mathcal H_{\text{int}}\!\to\!\mathcal H_B$ is therefore non-isometric, and its kernel should correspond to the ``null states'' of Ref.~\cite{Akers:2022qdl}. 
\end{enumerate}

%\section{Multi-partite entanglement}

So far we show that our state is well-approximated as Haar random state at the bipartite level. Finally, we briefly comment on the structure of multi-partite entanglement. Canonical examples of the AME state include stabilizer states that have a different multipartite entanglement structure compared to Haar random states~\cite{Li:2025nxv}. 

We can easily confirm that the present setup exhibits a non-vanishing Markov gap~\cite{Akers:2019gcv} based on holographic calculation of the reflected entropy~\cite{Dutta:2019gen}. This indicates the current AME states exhibit W-type tripartite entanglement that is not captured by random stabilizer states~\cite{Akers:2019gcv}. 
Therefore, our AME state is more likely to Haar random state than random stabilizer state. We leave more quantitative arguments for future work.

\section{Discussion}

We have argued that the quantum state dual to the special extremal slice is AME. It is worth noting that AME states are not expected to exist in arbitrary circumstances (see, \cite{scott2004multipartite,huber2017absolutely,huber2018bounds}). Our result provides an example of an AME state in a continuum field-theory setting. Since AME state can be approximated as a random state, this finding also helps to bridge the gap between the physics of black hole interior and quantum information theoretic toy models discussed in \cite{Akers:2022qdl}. Our results also provide a clear diagnostic for interior holography: dual theory must reproduce the AME property on the special extremal slice, offering a stringent benchmark for future approaches. It would be interesting to test it explicitly in the context of $T\overline{T}$-deformation and related approaches.

We have extracted the dimension of the black hole interior Hilbert space via the entanglement entropy. 
Our setup is similar to Cauchy slice holography \cite{Araujo-Regado:2022gvw}, in that we impose Dirichlet boundary conditions on a time slice. In such formulations the energy spectrum typically becomes complex in interior, implying that the dual theory is non-Hermitian. Consequently, the reduced density matrix is better regarded as a transition matrix than as a conventional density matrix, suggesting that the computed entropy should be viewed as the pseudo entropy~\cite{Nakata:2020luh}. Alternatively, one may view our setup as a particular deformation of timelike entanglement~\cite{Doi:2022iyj} which is again interpreted as the pseudo entropy.

Even if the quantity we compute must strictly be interpreted as pseudo entropy, our result can still be tied to the dimension of the interior Hilbert space, because holographic pseudo entropy is known not to suffer from the ``amplification’’ phenomenon in which its value exceeds the logarithm of dimension of the Hilbert space~\cite{Ishiyama:2022odv}. We also find no indication of unphysical behaviour such as violations of entropic inequalities within our setting.

Our conclusions are restricted to the pure gravity sector and the leading order of $G_N$ expansion. Including matter fields will require a more careful treatment of boundary conditions, akin to what is seen in $T\overline{T}$-deformed theories~\cite{Kraus:2018xrn,Hartman:2018tkw,Guica:2019nzm}; in particular, an appropriate choice of boundary condition may significantly alter the structure of the dual theory. Achieving a complete understanding of holography on interior screens with matter fields remains an important direction for future work.

%To our knowledge, this study provides the first example of an AME state in a continuum field-theory setting. 

\begin{acknowledgments}
\section{Acknowledgments}
We thank Kanato Goto, Shunichiro Kinoshita, Takato Mori, and Tomonori Ugajin for fruitful comments and discussions. T.A.~is supported by JSPS KAKENHI Grant No.~24K22886. K.T.~is supported by MEXT KAKENHI Grant No.~24H00972. K.T. would like to thank the organizers and participants of the conference ``Recent Developments in Black Holes and Quantum Gravity’’ at Yukawa Institute for Theoretical Physics and ``Quantum Connections: Linking Information, Gravity, and Many-Body Physics’’ where the part of this work was presented.  
\end{acknowledgments}

\bibliography{reference.bib}% Produces the bibliography via BibTeX.

\end{document}